\documentclass[twocolumn,noshowpacs,preprintnumbers,amsmath,amssymb]{revtex4-2}
\usepackage{graphicx}
\usepackage{dcolumn}
\usepackage{bm}
\usepackage{color}
\begin{document}

\title{
Resonant switching current detector 
based on underdamped Josephson junctions}

\author{Vladimir M. Krasnov}
\email{Vladimir.Krasnov@fysik.su.se}
\affiliation{Department of Physics, Stockholm University, AlbaNova University
Center, SE-10691 Stockholm, Sweden.}


\begin{abstract}
Current-biased Josephson junctions can act as detectors of electromagnetic radiation. At optimal conditions, their sensitivity is limited by fluctuations causing stochastic switching from the superconducting to the resistive state. This work provides a quantitative description of a stochastic switching current detector, based on an underdamped Josephson junction. It is shown that activation of a Josephson plasma resonance can greatly enhance the detector responsivity in proportion to the quality factor of the junction. The ways of tuning the detector for achieving optimal operation are discussed. For realistic parameters of Nb/AlOx/Nb tunnel junctions, the sensitivity 
can reach values of $S\simeq 5\times10^{12}$ (V/W). 

\end{abstract}

\maketitle

\section*{Introduction}

Sensitive, compact broad-range THz detectors are required for various applications, ranging from security and environmental monitoring, medical imaging, chemical analysis and fundamental research. Despite a large variety of available THz detectors (for review see e.g., \cite{Zhang_2016,Lewis_2019,Kasjoo_2020,Eisman_2011}), the search for a perfect sensor remains an active topic. The key parameter, which requires further improvement is the sensitivity, which is particularly important for passive THz detectors suitable for remote monitoring and imaging. 
The sensitivity can be greatly improved by cooling down to cryogenic temperatures, which both reduces the overall noise $\propto k_B T$ and removes the background black-body radiation. 
Ultra-sensitive cryogenic detectors based on semiconductors \cite{Astafiev_2000}, graphene \cite{Auton_2017,Kim_2020,Kim_2021} and superconductors \cite{Lau_2023} have been demonstrated. Superconductors are particularly well suited for THz detection because characteristics energy scales (Josephson energy, superconducting gap, e.t.c.) are naturally placed in the THz range and could be easily tuned for achieving optimal operation. 

A current-biased Josephson junction (JJ) demonstrates potential as a highly sensitive detector of electromagnetic waves, with the capability for single-photon resolution across a wide frequency range spanning from microwave (MW) to terahertz frequencies  \cite{Grimes_1968,Vernon_1972,Vystavkin_1974,Tucker_1985,Devyatov_1986,GrJensen_2004,McDermott_2012,Kuzmin_2013,Andersen_2013,Nakamura_2016,Ilichev_2017,Borodianskyi_2017,Kuzmin_2020, Divin_2020,Cattaneo_2021,Torrioli_2022}. 
The impacting electromagnetic wave induces a alternating current, $I_{THz}$, in the JJ. If bias current is slightly below the critical current, $I_c$, a very small $I_{THz}$ may cause switching of the junction from the superconducting to the resistive state. 
An underdamped JJ will stay in the resistive state until the bias current is reduced to a retrapping current, $I_r$ \cite{Kuzmin_2013,Ilichev_2017,Krasnov_2005,Krasnov_2007}. Such latching enables a long and large electric response to a small $I_{THz}$, 
thus enabling an easily measurable signal. The responsivity of such switching current detector (SCD) can be very high and is limited only by thermal and/or quantum fluctuations of phase 
\cite{Devyatov_1986,McDermott_2012,Kuzmin_2013,Andersen_2013,Nakamura_2016,Ilichev_2017,Kuzmin_2020,Torrioli_2022}. 
Therefore, stochastic effects associated with fluctuations, are essential for the analysis of SCD. 

The operation frequency range of SCD is determined by the Josephson plasma frequency, $\omega_p$ and the characteristic frequency, $f_c = (2e/h) V_c$, where $V_c = I_c R_n$ and $R_n$ is the normal junction resistance. For low-$T_c$ superconductors it is sub-THz, but for high-$T_c$ it is in the THz range \cite{Divin_2020,Borodianskyi_2017,Katterwe_2012,Cattaneo_2021}. Intrinsic Josephson junctions (IJJ's) in layered Bi$_2$Sr$_2$CaCu$_2$O$_{8+\delta}$ (Bi-2212) cuprates \cite{Kleiner_1994} have the highest $V_c \gtrsim 30$ mV \cite{Krasnov_2002}, facilitating operation in excess of 10 THz \cite{Borodianskyi_2017,Katterwe_2012}. The atomic scale of IJJ's enables a strong coupling between junctions and allows integration of a large number of IJJ's in a compact device 
\cite{Ozyuzer_2007,Benseman_2013,Kashiwagi_2015,HBWang_2019,Ono_2020,Kakeya_2020,Borodianskyi_2017,Cattaneo_2021}. Arrays with many JJs can be used for cascade multiplication of the readout signal \cite{Cybart_2019,Grebenchuk_2022,Cattaneo_2024} and for achieving impedance matching with open space, which is required for the effective absorption of incoming radiation \cite{MKrasnov_2021,Balanis,Krasnov_2023,Krasnov_2010}. 

In this work I present theoretical and numerical analysis of a stochastic SCD, based on an underdamped JJ. Although SCD is well known, to my knowledge there has been no comprehensive quantitative analysis of its operation. My goal is to fill this knowledge gap. As I will show, the performance of SCD is very sensitive to operation conditions. The key physical phenomenon that boosts SCD sensitivity is a resonant activation at the Josephson plasma frequency \cite{Larkin_1986,Devoret_1987,GrJensen_2004,Siddiqi_2005}. 
Various ways of tuning SCD for achieving optimal operation, are discussed. 
It is shown that the ultimate limit of sensitivity is determined by the phase diffusion phenomenon \cite{Kautz_1990,Pekola_2005,Krasnov_2007,Krasnov_2005}. Calculations based on realistic parameters for Nb/AlOx/Nb tunnel junctions show that the sensitivity 
at $T=1$~K can reach 
$S\simeq 5\times10^{12}$ (V/W). 

The paper is organized as follows. First we will recollect basic concepts on fluctuation-induced statistics in the absence of radiation and on MW response in the absence of fluctuations, followed by consideration of a general problem of MW response in the presence of fluctuations. We will start with the case of non-resonant escape at low frequencies and then consider resonant activation at $\omega_p$. In the end we will discuss limitations and ultimate performance. 

\section*{Results}

We consider an underdamped JJ with a quality factor, 
\begin{equation*}
  Q_0 = \omega_{p0} R_{QP} C \gg 1.
\end{equation*}
Here 
\begin{equation*}
  \omega_{p0}=\sqrt{\frac{2\pi}{\Phi_0}\frac{I_{c0}}{C}}
\end{equation*}
is the zero-bias plasma frequency, $I_{c0}$ is the fluctuation-free critical current, $\Phi_0$ is the flux quantum, $C$ is capacitance and $R_{QP}$ is the low-bias (subgap) quasiparticle (QP) resistance, which for tunnel JJs is much larger than the high-bias $R_n$. Photon detection by overdamped JJ with $Q_0\lesssim 1$ has been studied earlier \cite{Grimes_1968,Vernon_1972,Vystavkin_1974} (for an overview see e.g. Ch. 11.5 in Ref. \cite{Barone}). The sensitivity $S$
(V/W) of such the detector is proportional to the differential resistance of the current-voltage ($I$-$V$) characteristics at the bias point, $R_d = dV/dI(I)$. However, this simple description is inapplicable for us because underdamped JJs have an abrupt switching from the superconducting 
to the resistive 
state. For tunnel JJs the voltage jumps from zero to a large sum-gap value, $V_g=2\Delta/e$. 
In this case $R_d=\infty$ and the device operation can be described only statistically 
in terms of switching probabilities. 

\begin{figure*}[t]
    \centering
    \includegraphics[width=0.99\textwidth]{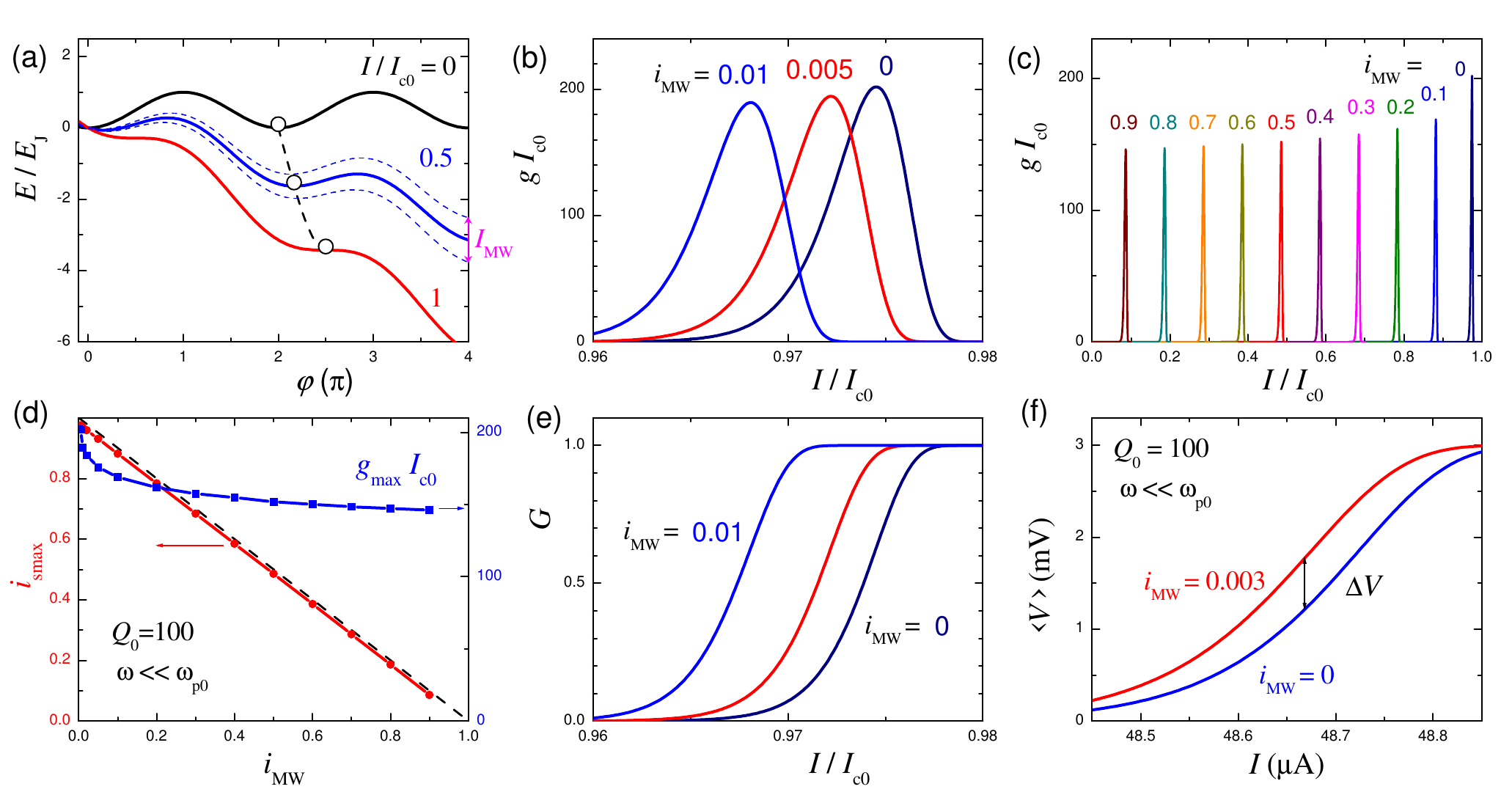}
    \caption{(Color online). Numerical simulation of non-resonant SCD operation at low frequencies, $\omega \ll \omega_p$. (a) Josephson wash-board potentials at three bias currents, $I/I_{c0}=0$, 0.5, and 1 (solid lines). Circles mark the stationary phase positions. Dashed blue lines illustrate shaking of potential by the microwave current, $I_{MW}$. (b) and (c) Calculated switching current histograms (b) at small MW amplitudes $I_{MW}/I_{c0}=0$, 0.005 (red) and 0.01 (blue), and (c) at large $I_{MW}/I_{c0}$. (d) The most probable switching current, $i_{smax}$ (red circles, left axis), and the corresponding maximum switching probability density, $g_{max}$ (blue squares, right axis), as a function of MW amplitude $i_{MW}$. The dashed black line shows the fluctuation-free $I_c$ according to Eq. (5). (e) Switching probabilities for the same $i_{MW}$ as in (b). (f) The time-average current-voltage characteristics of the junction without MW (blue) and with $i_{MW} =0.003$ (red). The vertical shift of the curves, $\Delta V$, represents the SCD response at a given bias current.}
    \label{fig:fig1}
\end{figure*}

Dynamics of a JJ 
is equivalent to motion of a particle in a tilted washboard potential \cite{Barone}, $U(\varphi)=E_{J0} [1-\cos\varphi - i\varphi ]$, as sketched in Fig. \ref{fig:fig1} (a).
Here $\varphi$ is the Josephson phase difference, $E_{J0}=(\Phi_0/2\pi) I_{c0}$ is the Josephson energy and $i=I/I_{c0}$. The washboard has bias-dependent parameters: the barrier height, $\Delta U = 2 E_{J0} [ (1-i^2)^{1/2} -i\text{arccos}(i) ]$,
the eigenfrequency $ \omega_p \simeq \omega_{p0} (1-i^2 )^{1/4}$,
and the quality factor, $Q=\omega_p R_{QP} C$.
Below I will show calculations for the typical experimental situation of ac-bias, $I(t)=I_{b} \sin (\omega_b t)$, at $\omega_b/2\pi=150$ Hz, $T=1$ K, and parameters corresponding to Nb/AlOx/Nb tunnel junctions with $V_c=1$ mV and $V_g=3$ mV \cite{Koshelets_2019}. 

\subsection*{Switching statistics without external radiation}

Fluctuations cause premature escape out of the well at a switching current $I_s<I_{c0}$. The escape rate in the absence of radiation can be written as \cite{Martinis_1987,Grabert_1987,Martinis_1988}. 
\begin{equation}
    \Gamma_0(I) = a(I)\frac{\omega_p(I)}{2\pi}\exp{\left[ -\frac{\Delta U(I)}{k_b T} \right]}.
    \label{G0} 
\end{equation}
For underdamped JJs excited by thermal fluctuations, the prefactor can be written as \cite{Buttiker_1982}, $a(I) \simeq 4[(1+Q(I) k_b T/1.8 \Delta U(I))^{1/2}+1]^{-2}$. The effect of quantum fluctuations at $T\rightarrow 0$ can be easily taken into account by introducing an effective escape temperature $T_{esc}\sim \hbar\omega_{p}/2\pi k_B$ \cite{Martinis_1987}. 

The probability density for switching in the bias interval $I_b < I < I_b + dI$ is 
\begin{equation}
    g (I) = \frac{\Gamma (I)}{dI/dt}\left[1- G(I) \right],
    \label{P0dens}
\end{equation}
where $dI/dt$ is the bias ramp rate and  
\begin{equation}
    G(I)=\int_0^{I}{g(I) dI}
    \label{P0}
\end{equation}
is the total probability of switching upon ramping up to current $I$. Eqs. (\ref{P0dens}) and (\ref{P0}) form a recurrent equation, which can be easily solved numerically. 

\subsection*{Microwave response without fluctuations}

In the absence of fluctuations the MW response of JJ has been studied in connection with the analysis of Shapiro steps. Electromagnetic waves induce both voltage and current oscillations in the junction. Since JJ is essentially non-linear, a self-consistent treatment of both voltage and current components is a difficult task. To simplify the analysis, voltage- or current-source approximations are employed \cite{Barone}. 

\begin{figure*}[t]
    \centering
    \includegraphics[width=0.99\textwidth]{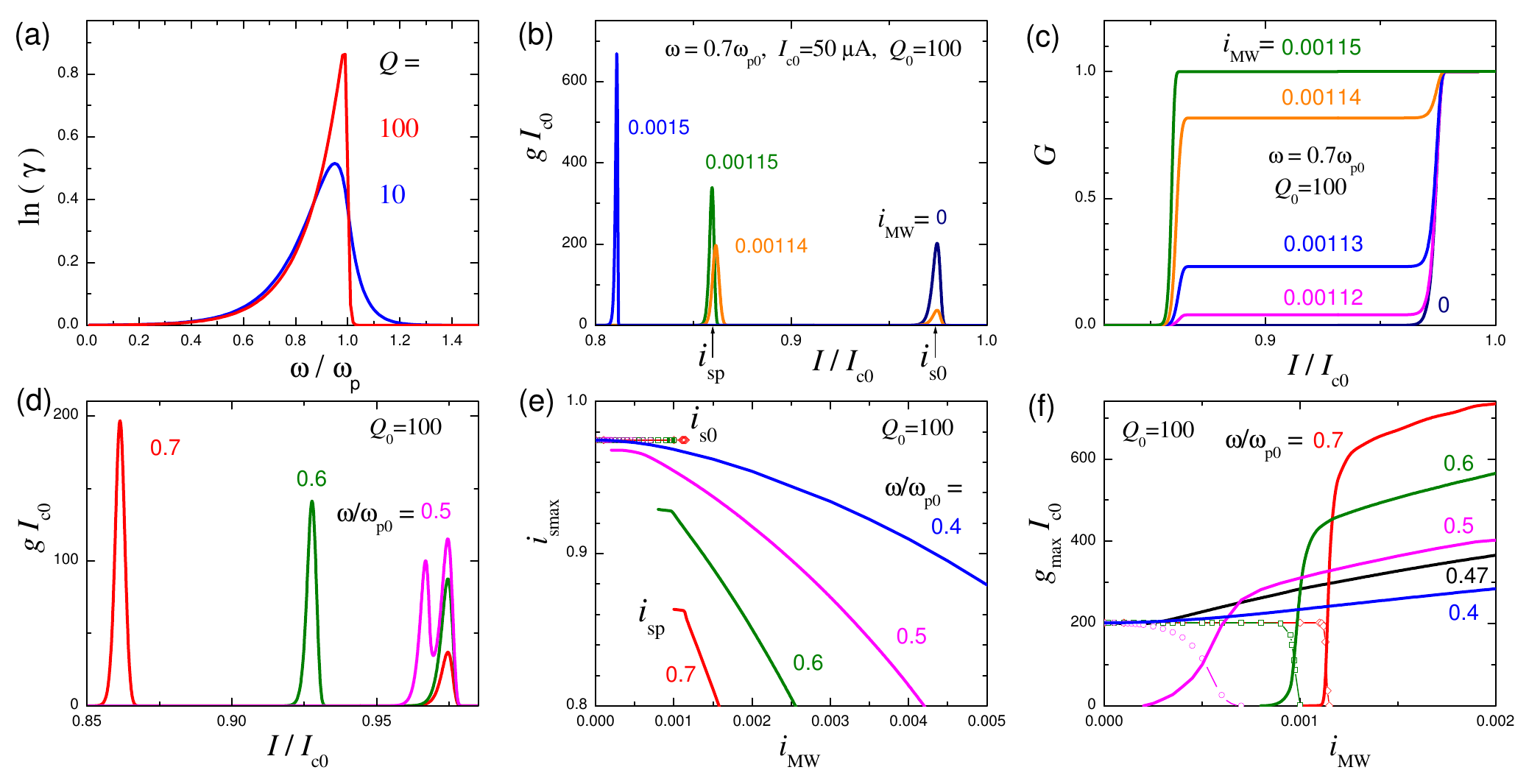}
    \caption{(Color online). Operation of a resonant SCD at $\omega \sim \omega_p$. (a) Resonant escape rate gains, $\gamma(\omega)$, for quality factors $Q=10$ (blue) and 100 (red). (b) Switching current histograms for different MW amplitudes at $\omega=0.7 \omega_{p0}$ and $Q_0=100$. Splitting of the histogram at $i_{MW}=0.00114$ (orange) is caused by the excitation of the Josephson plasma resonance. (c) Switching probabilities in the bifurcation region for $\omega =0.7~\omega_{p0}$. The left and right steps at $i_{sp}$ and $i_{s0}$ correspond to resonant and non-resonant escapes, respectively. (d) Split-histograms in the bifurcation region for different frequencies, $\omega/\omega_{p0}=0.5$ (magenta), 0.6 (olive) and 0.7 (red). (e) and (f) MW amplitude-dependence of (e) the most probable switching current, $i_{smax}$, and (f) the corresponding maxima of the probability density, $g_{max}$, for different frequencies. Symbols and lines represent non-resonant, $i_{s0}$, and resonant, $i_{sp}$, peaks, correspondingly. The threshold (gap) between two peaks disappears at $\omega/\omega_{p0}\lesssim 0.47$. }
    \label{fig:fig2}
\end{figure*}

The simplest is the voltage-source approximation. It postulates the harmonic MW voltage, $V_{MW}\cos(\omega t)$ and yields an explicit expression for the fluctuation-free critical current \cite{Barone}:
\begin{equation}
    I_{c} = I_{c0} J_0 \left[\frac{2eV_{MW}}{\hbar\omega}\right],
    \label{I0Bessel}
\end{equation}
where $J_0$ is the zero-order Bessel function. Unfortunately, this simple expression is not applicable for the analysis of SCD in the most interesting frequency range $\omega<\omega_{p0}$. Eq. (\ref{I0Bessel}) works only at frequencies well above $\omega_{p0}$, but at low frequencies it gives a qualitatively incorrect prediction of increasing (diverging) responsivity with decreasing $\omega$, which is opposite to reality (see Fig. \ref{fig:fig2}(e) below). 

The current-source approximation postulates a harmonic MW current, $I_{MW}\sin(\omega t)$. This is a more complex approach, which allows only numerical solution. However, it provides a physically correct results at low frequencies. In agreement with experiment, it predicts a linear reduction of $I_{c}$ vs. $I_{MW}$ at $\omega\ll\omega_{p0}$ (see ch. 11.3 in Ref. \cite{Barone}). This has a clear physical explanation. At low frequencies the junction dynamics is quasi stationary. Therefore, bias and MW currents simply add up. The switching occurs when the total current 
exceeds $I_{c0}$, which yields, 
\begin{equation}
    I_{c}\simeq I_{c0}-I_{MW}~~~~(\omega\ll\omega_{p0}).
    \label{I0lin}
\end{equation}

At frequencies well above $\omega_{p0}$, current- and voltage-source solutions are in a qualitative agreement with each other \cite{Barone}, although the connection between $V_{MW}$, $I_{MW}$ and the incoming MW power, $P_{in}$, remains uncertain. To couple them, first we need to introduce a high-frequency impedance $Z_{MW}$ of the detector, so that $V_{MW}=Z_{MW}I_{MW}$. The absorbed MW power is then 
\begin{equation}
    P_{a} =\frac{I_{MW}^2R_{MW}}{2},
    \label{Pa}
\end{equation}
where the MW resistance, $R_{MW}=\text{Re}(Z_{MW})$. This resistance has nothing to do with $R_{QP}$. For example, in the current-source model, vertical shaking of the potential by $I_{MW}$ leads also to a horizontal motion of the particle, as indicated by the black dashed line in Fig. \ref{fig:fig1} (a). The associated MW voltage is connected to $I_{MW}$ via the impedance of the nonlinear Josephson induction \cite{Barone}. In reality, $Z_{MW}$ represents the total 
(environmental)
MW impedance of the device, including electrodes, which should be designed and act as antennas for effective catching of MW \cite{Krasnov_2023}. 
Second, we need to introduce the absorption efficiency,
\begin{equation}
    \chi=\frac{P_a}{P_{in}}.
    \label{chi}
\end{equation}
The maximum theoretical value of $\chi$ is 0.5, achieved at the impedance matching condition \cite{Krasnov_2010}. However, in reality $\chi$ is somewhat lower because of the QP damping \cite{Krasnov_2023}, which causes a leakage current and an additional power consumption on heat production. For a quasi-optical supply of radiation, $\chi$ is further reduced in proportion to the ration of the effective absorption area of the detector antenna to the beam spot area, which can be much smaller than unity if the size of the device is smaller than $\lambda_0$. Thus, 
\begin{equation}
    P_{in} =\frac{I_{MW}^2R_{MW}}{2\chi}.
    \label{Pin}
\end{equation}
The values of $R_{MW}$ and $\chi$ are often ill-defined. In what follows I shall assume $R_{MW}=100~\Omega$ \cite{Siddiqi_2005} and $\chi=0.5$, assuming the best case scenario of a well-matched and optimized detector \cite{Krasnov_2010}.

\subsection*{Microwave response with fluctuations}

In the presence of MW, the total current is oscillating,
\begin{equation}
    I_{tot}(t)= I(t)+I_{MW}\sin(\omega t),
    \label{Ib_curr}
\end{equation}
where
\begin{equation*}
    I(t)=I_b\sin(\omega_b t).
\end{equation*}
This leads to shaking of the washboard potential, as shown in Fig. \ref{fig:fig1} (a), which enhances the escape rate,
\begin{equation}
    \Gamma_{MW}=\gamma\Gamma_0,
    \label{gamma}
\end{equation}
by some gain factor $\gamma$. The most prominent enhancement occurs when $\omega$ coincides with the eigenfrequency $\omega_p$, leading to excitation of a plasma resonance. 

\subsubsection*{Non-resonant escape at $\omega \ll \omega_p$}

In the low-frequency limit, the dynamics is quasi-stationary. Therefore, barrier characteristics $\omega_p(I_{tot})$ and $\Delta U(I_{tot})$ are well defined at every time instance and the escape rate is obtained from Eq. (\ref{G0}) with a given time-dependent $I_{tot}(t)$, Eq. (\ref{Ib_curr}). Thus the modified escape rate is obtained by time-averaging over the MW period, $\tau=2\pi/\omega$,
\begin{equation}
    \Gamma_{MW}(I) = \frac{1}{\tau}\int_0^{\tau}{\Gamma_0[I_{tot}(t)]dt}.
    \label{Gmw}
\end{equation}
Switching probabilities are obtained by substituting it in Eqs. (\ref{P0dens}) and (\ref{P0}), taking into account the actual time dependent bias ramp rate,
which in our case is, $dI/dt=I_b\omega_b\cos( \omega_b t)$. Note that the escape gain $\gamma$ in this limit is frequency independent.  

\begin{figure}[t]
    \centering
    \includegraphics[width=0.5\textwidth]{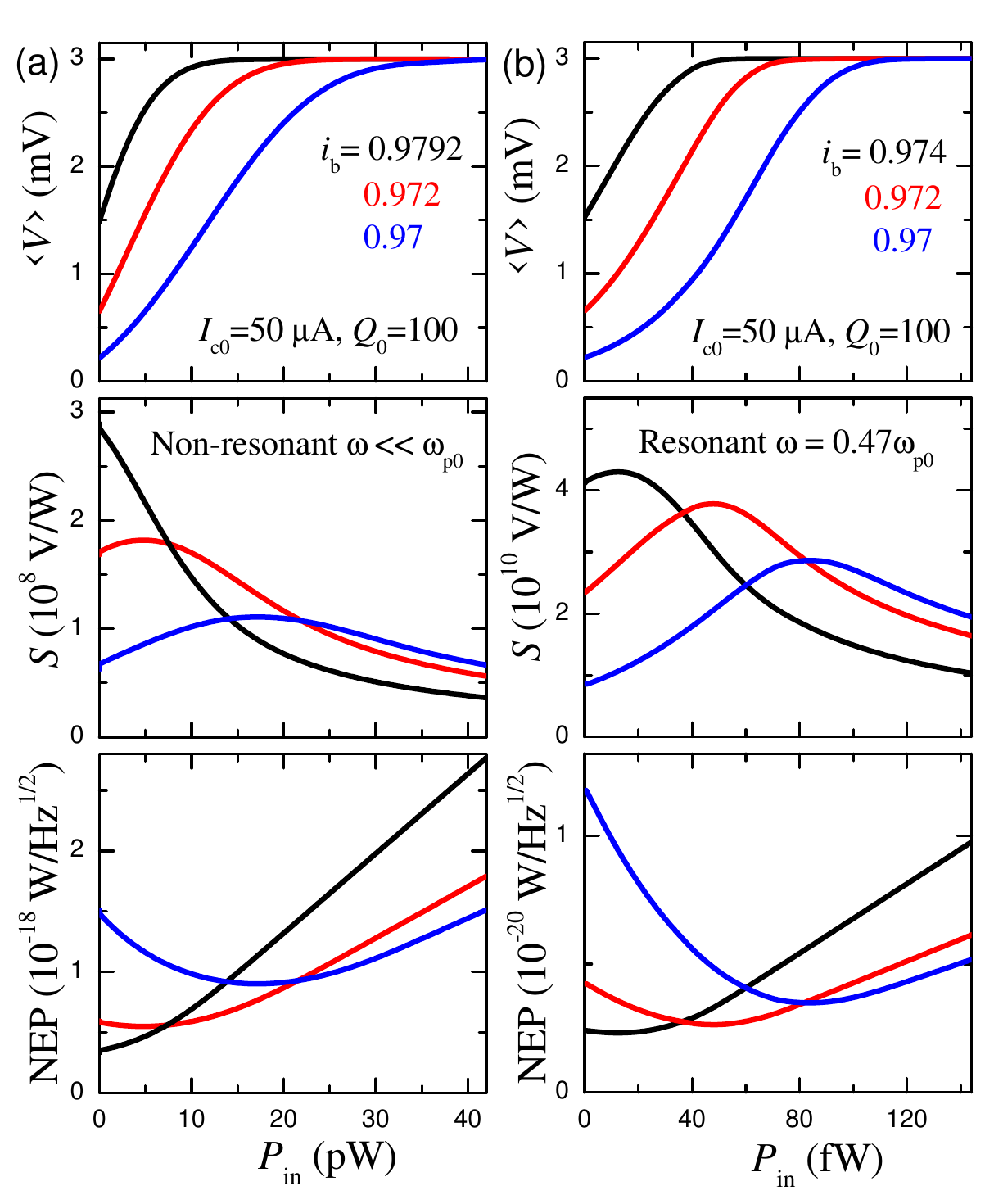}
    \caption{(Color online). Bias dependence of (a) non-resonant SCD at $\omega \ll\omega_{p0}$ and (b) resonant SCD at the optimal frequency, $\omega = 0.47 \omega_{p0}$. Top panels show time-average voltages as a function of incoming MW power for three bias current amplitudes. Middle and bottom panels show corresponding sensitivities and noise-equivalent powers, respectively. [Erratum: NEP values are incorrect \cite{Erratum_2025}.] }
    \label{fig:fig3}
\end{figure}

Figs. \ref{fig:fig1} (b) and (c) show thus calculated probability densities for different MW amplitudes, $i_{MW}=I_{MW}/I_{c0}$, for a JJ with $I_{c0}=50~\mu$A. They correspond to experimentally measurable switching current histograms, $I_s(I)$. It is seen that the MW excitation leads to a shift of histograms with successive reduction of the most probable switching current $I_{smax}$. Fig. \ref{fig:fig1} (d) represents the MW-amplitude dependence of $i_{smax}=I_{smax}/I_{c0}$ (red squares, left axis). The black dashed line shows the fluctuation free $I_c$ according to Eq. (\ref{I0lin}). 
Blue circles (right axis) represent the maximum probability density $g_{max}$. It exhibits a rapid decline at small $i_{MW}$ and then saturates at larger $i_{MW}$. As will be shown below, the rapid decline ensures a non-diverging responsivity at $i_{MW}\rightarrow 0$, see Fig. \ref{fig:fig3} (a).
Fig. \ref{fig:fig1} (e) shows switching probabilities $G(i)$ for the same $i_{MW}$ as in Fig. \ref{fig:fig1} (b). Since tunnel junctions switch to the bias-independent gap voltage, $G$ represent the time-average dc-voltage, $\langle V \rangle = G V_g$. Note that although $\langle V \rangle$ varies gradually with $I_b$, the switching remains abrupt at each measurement. 

Fig. \ref{fig:fig1} (f) clarifies the operation principle of the SCD. Here time-average $V$-$I$ characteristics, $\langle V(I) \rangle = G(I) V_g$, are shown without MW (blue) and with a small $i_{MW}=0.003$ (red). The voltage response, $\Delta V$, 
is equal to the vertical shift of the curves. At small $i_{MW}$, it is proportional to the shift and the slope of the step-like $G(I)$, 
\begin{equation}
    \Delta V \simeq -V_g \frac{\partial G}{\partial i_s} \frac{\partial i_s}{\partial i_{MW}} i_{MW}.
\end{equation}
Therefore, maximum sensitivity is achieved at the steepest part in the middle of the step, $G \simeq 0.5$.

\subsubsection*{Resonant activation at $\omega \sim \omega_p$}


Josephson plasma resonance occurs when $\omega \sim \omega_p$ \cite{Larkin_1986,Devoret_1987,GrJensen_2004,Siddiqi_2005}. Since the amplitude of forced oscillations is proportional to $Q$, resonant activation greatly enhances the escape rate in underdamped junctions with $Q\gg 1$. In Ref. \cite{Devoret_1987} a fitting function was suggested for the resonant gain factor $\gamma$: 
\begin{eqnarray}
    \ln \gamma \simeq \frac{5 E_{J0}\Delta U}{(k_B T)^2} \frac{i_{MW}^2 Q}{(\omega_p/\omega_{p0})^2}  f(x),~~~x=\frac{\omega}{\omega_p}-1, ~~~~~~~~~\\
\begin{aligned}   
    f(x<0)= Q\left[\frac{e^{9x}}{2Q+9}\left(1-2x+\frac{2}{2Q+9}\right)\right.~~~~~~~~~~~~~~~~\\ 
    \left. +\frac{e^{2Qx}-e^{9x}}{9-2Q}\left(1+\frac{2}{9-2Q}\right) +\frac{2x e^{9x}}{9-2Q} \right],~~~~~~
\end{aligned}\\
    f(x>0)=Q e^{-2Qx}\left[\frac{1}{9+2Q}+\frac{2}{(9+2Q)^2} \right],~~~~~~~~~~~~~~~
\end{eqnarray}
(in these expressions I used explicit values for parameters $u=9$, $v=-2$ in Eq. (6.5) from Ref. \cite{Devoret_1987}, as suggested by the authors, and $c=5$, $\lambda=2Q$ as follows from Fig. 19 from Ref. \cite{Devoret_1987}).

Figure \ref{fig:fig2} (a) shows resonant escape gain factors for $Q=10$ and 100, calculated from Eqs. (13-15). They resemble response functions of a driven oscillator, but have a sharp exponential cutoff at $\omega>\omega_p$ \cite{Larkin_1986}. 
The resonant escape is easily analyzed by substituting Eqs. (13-15) in Eq. (10) and subsequently in Eqs. (2) and (3). 

Fig. \ref{fig:fig2} (b) shows calculated switching histograms for $\omega=0.7~\omega_{p0}$, $I_{c0}=50~\mu$A, $Q_0=100$, $T=1$ K for several MW amplitudes, $i_{MW}$.
Plasma resonance is manifested by a bifurcation at some threshold MW amplitude, leading to splitting of histograms \cite{Devoret_1987,GrJensen_2004,Ilichev_2017,Siddiqi_2005}. An additional plasma peak emerges at a switching current, $i_{sp}$, lower than that in the absence of MW, $i_{s0}$.  
Fig. \ref{fig:fig2} (c) shows the two-step switching probabilities in the bifurcation region.

The plasma peak corresponds to the resonant condition $\omega  \simeq \omega_p(i_{sp})$, yielding,
\begin{equation}
i_{sp} \simeq \sqrt{1-(\omega/\omega_{p0})^4}.
\label{i_sp}
\end{equation}
Fig. \ref{fig:fig2} (d) shows split-histograms at different MW frequencies. It can be seen that $i_{sp}$ moves upwards with reducing $\omega$, in accord with Eq. (16). Figs. \ref{fig:fig2} (e) and (f) show MW amplitude dependencies of (e) peak currents and (d) heights, $g_{max}$, for several MW frequencies. From Fig. \ref{fig:fig2} (e) it is seen that $i_{smax}$ decays approximately parabolically with $i_{MW}$. The $i_{smax}(i_{MW})$ dependency becomes stronger with increasing $\omega$, which is in contrast to Eq. (4), but in line with Eq. (16).  

The bifurcation region can be extremely narrow: in Fig. \ref{fig:fig2} (c) it starts at $i_{MW}\simeq 0.00112$ and finishes at 0.00115. Its width, $\delta I_{MW} \simeq 3\times 10^{-5}I_{c_0}$, is two orders of magnitude narrower than that in the absence of MW, see Fig. \ref{fig:fig1} (b). A very small $\delta I_{MW}$ implies a very high sensitivity to MW. However, the problem for detector application is in the presence of a threshold $i_{MW}$, needed for reaching the bifurcation point, below which the response is much smaller. This can be seen from the almost constant sub-threshold $I_{smax}$ and $g_{max}$, as shown by open symbols in Figs. \ref{fig:fig2} (e) and (f). 
From Figs. \ref{fig:fig2} (d-f) it can be seen that the threshold is reduced with reducing frequency and disappears when the two peaks merge, 
\begin{equation}
  i_{sp}=i_{s0}.  
  \label{Optimum}
\end{equation}
This is the optimal operation condition for SCD. At this point the plasma resonance occurs within the accessible bias range for $i_{MW}\rightarrow 0$. For the chosen parameters in Fig. \ref{fig:fig2}, this occurs at $\omega=0.47~\omega_{p0}$, indicated by the black line in Fig. \ref{fig:fig2} (f). Inverting Eq. (16), we can see that SCD has optimal sensitivity at 
\begin{equation}
\omega^* =  \omega_{p0}[1-(i_{s0})^2]^{1/4}.    
\end{equation}
At higher frequencies the bifurcation threshold appears. At lower frequencies switching occurs before reaching the resonant conditions. 


\section*{Discussion}

A good sensor should have a large sensitivity, $ S = \Delta V / P_{in}$ (V/W) and a low $NEP$ (W/Hz$^{1/2}$). They are related to each other via the voltage noise, $S_V$ (V/Hz$^{1/2}$),
\begin{equation*}
    NEP=S_V/S.
\end{equation*}
However, equally important, it should have a high absorption efficiency, i.e., the ratio of absorbed to impacting MW power, $\chi$, in Eq. (\ref{chi}). A junction alone can not absorb the MW power because its size is much smaller than the wavelength, $\lambda_0$, \cite{Krasnov_2023}. Therefore, achieving a high efficiency requires implementation of impedance-matching MW antennas \cite{Krasnov_2023,MKrasnov_2021,Balanis}. 
In what follows I shall assume $R_{MW}=100~\Omega$ and  $\chi=0.5$, typical for well-matched MW devices.

\begin{figure*}[t]
    \centering
    \includegraphics[width=0.99\textwidth]{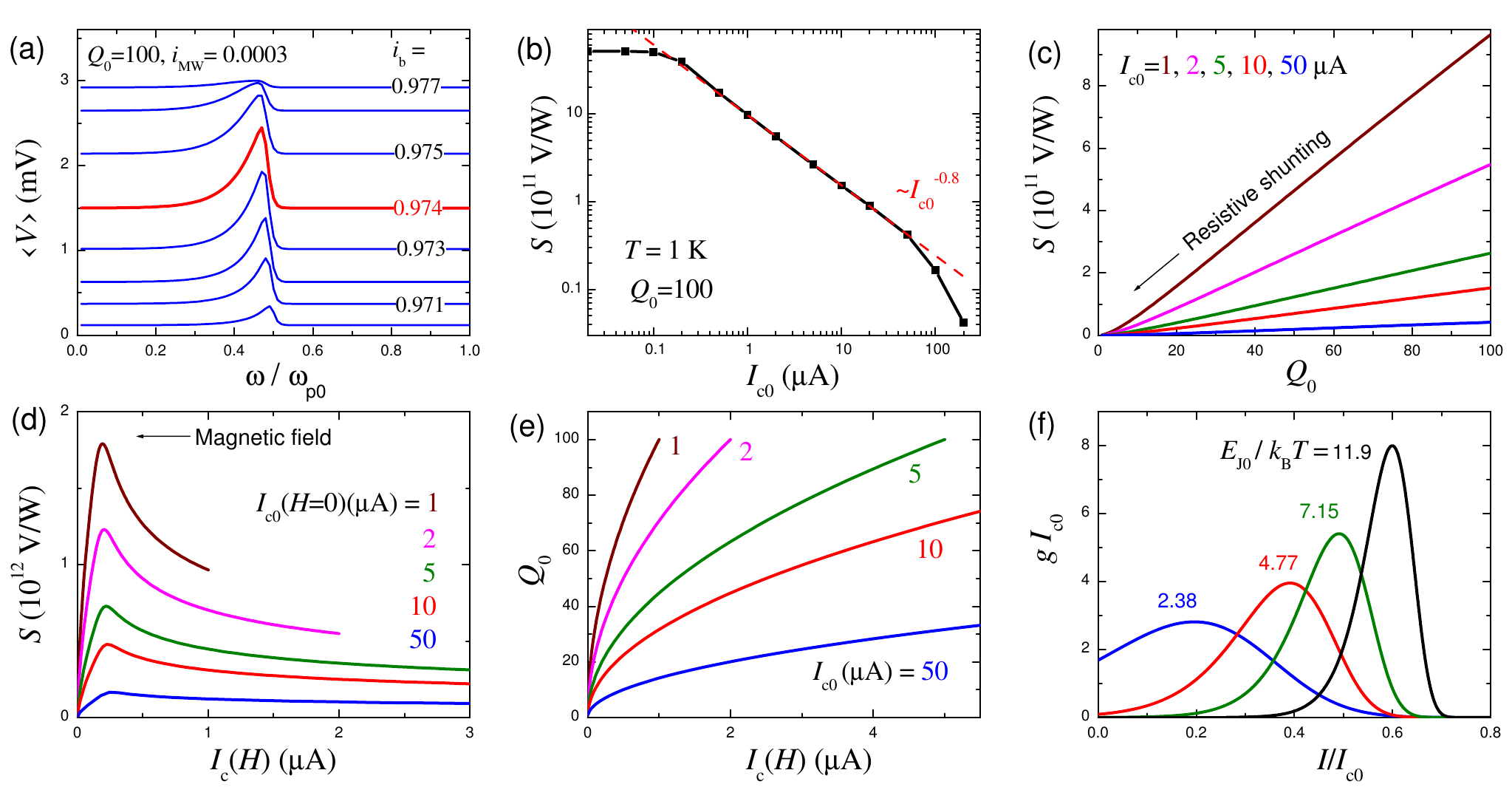}
    \caption{(Color online). Analysis of optimal operation and ultimate resolution of a resonant SCD. (a) Frequency dependence of the time-average voltage for different bias amplitudes and $i_{MW}=0.0003$. The peak at $\omega/\omega_{p0}\simeq 0.47$ is caused by the plasma resonance. (b) Effect of junction area variation. Optimal sensitivity is shown as a function of critical current, $I_{c0}$, for constant $Q_0=100$. The dashed line indicates that the sensitivity increases as $I_{c0}^{-0.8}$. (c) Effect of resistive shunting. Optimal sensitivities as a function of quality factor are shown for junctions with different area and $I_{c0}$. (d) and (e) Effect of magnetic field suppression of $I_c$. (d) Optimal sensitivity as a function of field-dependent $I_c(H)$ for junctions with different area and $I_{c0}$. (e) Corresponding variation of the quality factor. (f) Switching histograms for junctions with very small $I_{c0} = 0.1$ (blue), 0.2 (red), 0.3 (olive) and 0.5 (black) $\mu$A without MW radiation. The non-zero switching probability density at $I=0$ corresponds to the phase-diffusion regime of the junction.}  
    \label{fig:fig4}
\end{figure*}

Figure \ref{fig:fig3} summarizes performances of (a) non-resonant SCD at low frequencies and (b) resonant SCD at the optimal frequency, $\omega=0.47~\omega_{p0}$. They are based on simulations from Figs. \ref{fig:fig1} and \ref{fig:fig2}, respectively. Top panels show time-average voltage versus $P_{in}$ for three bias amplitudes, $i_b=I_b/I_{c0}$, close to $i_{s0}$. 
Middle panels represent sensitivities. The largest sensitivities at low $P_{in}$ are achieved for black curves at bias current corresponding to the middle of the steps $G(i,i_{MW}=0) = 0.5$. 
With increasing $P_{in}$ the response saturates at $V=V_g$, the voltage response becomes independent of $P_{in}$, the sensitivity decays as $1/P_{in}$ and NEP increases $\propto P_{in}$, as can be seen from middle and bottom panels in Fig. \ref{fig:fig3}. The saturation occurs when the shift of $i_{smax}$ exceeds the width of the histogram $g(i, i_{MW}=0)$, as for the blue curve at $i_{MW}=0.01$ from Fig. \ref{fig:fig1} (e). By reducing the bias current below the optimal point it is possible to extend the dynamic range at the expense of lower sensitivity. 
Bottom panel show NEP. Calculation of NEP requires estimation of the voltage noise $S_V$ in the measurement setup. It is in the range of few nV/Hz$^{1/2}$ for conventional dc-measurements, but could be greatly reduced by lock-in measurements \cite{Hovhannisyan_2021}. 
[Erratum: this assumption is incorrect, the noise of SCD is dominated by the much larger intrinsic telegraph noise \cite{Erratum_2025}.] 

Although the behavior of non-resonant and resonant SCD's from Figs. \ref{fig:fig3} (a) and (b) is qualitatively similar, there are important differences. First of all, the resonant SCD has a significantly higher sensitivity as a consequence of resonant activation. Another important difference is the frequency and $Q$-dependencies. The response of non-resonant SCD is frequency independent within the range of applicability, $\omega \ll \omega_p$. It is also almost independent of $Q$, which appears only in the prefactor $a$ of Eq. (\ref{G0}). For the resonant SCD the situation is completely different, as illustrated in Figure \ref{fig:fig4}.

Fig. \ref{fig:fig4} (a) shows frequency dependence of the time-average voltage 
at different bias amplitudes for a small $i_{MW}=0.0003$. Junction parameters are the same as in Fig. \ref{fig:fig2}. The response shows a sharp peak at $\omega=0.47 ~\omega_{p0}$, which corresponds to the plasma resonance in this bias range. The shape of the peak reflects the resonant gain factor, shown by the red line in Fig. \ref{fig:fig2} (a). The maximum response is obtained at $i_b$, corresponding to the middle of the switching step, $G(i_b,i_{MW}=0)=0.5$ (red curve). 
The optimal sensitivity at this bias is shown by black lines in Fig. \ref{fig:fig3} (b). Thus, the resonant SCD has a very strong frequency dependence. 

\subsection*{The ultimate sensitivity of SCD}

The sensitivity and NEP, shown in Fig. \ref{fig:fig3} (b) are remarkably good, in line or even better than the most sensitive photon detectors reported so far \cite{Lewis_2019,Auton_2017,Astafiev_2000,Kim_2020,Kim_2021}. However, this is not the ultimate limit of a resonant SCD. These calculations are made for fairly modest junction parameters and there are several ways to improve the sensitivity. Firstly, the responsivity is determined by the relative MW current, $i_{MW}=I_{MW}/I_{c0}$. Therefore, the sensitivity is increasing with decreasing $I_{c0}$. This can be done by reducing the junction area or by suppression of the critical current with magnetic field, $I_c(H)$. Secondly, the resonant gain is increasing with the quality factor, as seen from Fig. \ref{fig:fig2} (a). 
Thus, optimization of sensitivity requires proper tuning of junction parameters. 

Table I lists possible tuning parameters for optimization of SCD. The aim of tuning is to remove the bifurcation threshold by aligning the plasma resonance with the center of the switching histogram in the absence of radiation, Eq. (\ref{Optimum}).

Reducing the junction area, $A$, proportionally reduces $I_{c0}$
while keeping $Q_0$ and $\omega_{p0}$ unchanged. Indeed, the critical current density, $I_{c0}/A$, the capacitance per area $C/A$ and normal resistance per area $R_{QP}A$ remain constant. Therefore, $\omega_{p0} \propto \sqrt{I_{c0}/C}$ and $Q_0 \propto R_{QP}C$ are independent of JJ size. 
In Fig. \ref{fig:fig4} (b) we show a log-log plot of the optimal sensitivity for junctions with different area and the same $Q_0=100$. The red line indicates that $S$ scales as $I_{c0}^{-0.8}$, before it saturates at $S\simeq 5\times 10^{12}$ (V/W). 

The quality factor can be tuned by resistive shunting of the junction. In this case the $I_{c0}$ stays constant, but $Q_0$ is reduced. Fig. \ref{fig:fig4} (c) shows optimal sensitivities versus the quality factor for junctions with several fixed $I_{c0}$. It is seen that $S$ increases approximately linear with $Q_0$ at $Q_0\gg 1$, as expected for forced plasma oscillations. 

\begin{table*}[t]
    \centering
    \begin{tabular}{|l|c|c|c|c|c|}
    \hline
    
     ~Tunable parameters & $I_{c0}$ & $C$ & $R_{QP}$ & $\omega_{p0}$ & $Q_0$ \\
     \hline
     ~Critical current density, $J_{c0}$~ & $\propto J_{c0}$ &  ~slightly~ & $\propto 1/J_{c0}$ & $\propto J_{c0}^{1/2}$ & $\propto J_{c0}^{-1/2}$\\
     ~Junction area, $A$ & $\propto A$ & $\propto A$ & $\propto 1/A$ & ~negligible~ & ~~negligible~~ \\
     ~Resistive shunting, $R$ & ~negligible~ & negligible & $\propto R$ & negligible & $\propto R$ \\
     ~Capacitive shunting, $C$ & negligible & $\propto C$ & ~negligible~ & $\propto C^{-1/2}$ & $\propto C^{1/2}$\\
     ~Magnetic field, $H$ & $I_c(H)$ & negligible & negligible & $\propto I_{c}^{1/2}$ & $\propto I_{c}^{1/2}$ \\
     ~Temperature, $T$ & $I_{c0}(T)$ & negligible & $R_{QP} (T)$ & $\propto I_{c0}^{1/2}$ & $\propto I_{c0}^{1/2}R_{QP}$ \\
     \hline
    \end{tabular}
    \caption{Tunning parameters for optimization of SCD and their influence on junction characteristics. The aim of tuning is to remove the bifurcation threshold by aligning the plasma resonance with the center of the switching histogram in the absence of radiation. In the first line the critical current density is changed by varying the oxide thickness. This does slightly affect the capacitance, but not as much as $J_{c0}$. In this case the characteristic voltage $V_c$ remains approximately constant, so that $R_{QP}=V_c/I_{c0} \propto 1/J_{c0}$.} 
    \label{tab:1}
\end{table*}

Finally, it is possible to in-situ tune junction parameters by changing temperature or applying magnetic field \cite{Martinis_1987}. Figs. \ref{fig:fig4} (d) and (e) illustrate the effect of magnetic field. Here we consider junctions with different areas having the same $Q_0$, but different $I_{c0}$ at $H=0$. With application of magnetic field $I_c(H)$ is suppressed and $Q_0$ is reduced $\propto I_{c0}^{1/2}$, as shown in Fig. \ref{fig:fig4} (e). This drives the sensitivity in opposite directions and leads to appearance of maxima in $S$, as shown in Fig. \ref{fig:fig4} (d). 

Yet the ultimate sensitivity is limited by another phenomenon - entrance in the phase-diffusion state \cite{Kautz_1990,Pekola_2005,Krasnov_2005,Krasnov_2007}. With reducing $I_{c0}$, the Josephson energy is reducing proportionally. When $E_{J0}$ becomes comparable with $k_B T$, the particle is no longer stationary in the washboard, but can diffuse in the potential even without a tilt at $I_b=0$ \cite{Siddiqi_2005}. Fig. \ref{fig:fig4} (f) shows switching histograms without MW for $I_{c0}=0.1$, 0.2, 0.3 and 0.5 $\mu$A, corresponding to $E_{J0}/k_B T \simeq$ 2.38, 4.22, 7.15 and 11.9, respectively. It is seen that $g(I=0)$ becomes significantly larger than zero at $I_{c0} = 0.2~\mu$A (red curve) for $T=1$ K. From Figs. \ref{fig:fig4} (b) and (d) it can be seen that the maxima of sensitivity are reached at the same current $I_{c0}\sim 0.2~\mu$A, irrespective of $Q_0$.  
As seen from Fig. \ref{fig:fig4} (f), entering in the phase-diffusion state broadens the switching histograms, which deteriorates the sensor performance. Nevertheless, the sensor may still work even in the phase diffusion state. Moreover, sharpening of switching histograms may occur in moderately damped junctions \cite{Pekola_2005,Krasnov_2005,Krasnov_2007}, but the analysis of operation at the edge of phase-diffusion requires proper analysis of the retrapping process, which is beyond the scope of this work. 

The phase diffusion can be reduced by lowering the temperature, with the ultimate limit set by the crossover temperature to macroscopic quantum tunneling (MQT) $T_{MQT}\simeq \hbar \omega_p/2\pi k_B$, which is typically few tens of mK \cite{Martinis_1987,Grabert_1987,Martinis_1988,Devyatov_1986}. 
The sensitivity can be improved by increasing the quality factor, e.g. by means of a capacitive shunting of the junction \cite{Krasnov_2005,Siddiqi_2005}. However, this is accompanied by the reduction of $\omega_{p0}\propto C^{-1/2}$, reducing the frequency range. Cascade amplification of the readout voltage using arrays of coupled JJs can also strongly enhance the ultimate sensitivity \cite{Cattaneo_2024}.

To check how realistic are the obtained numbers, I make a comparison with experimental data by Siddiqi et.al. \cite{Siddiqi_2005}, who have studied  a ``pure” plasma resonance in an unbiased JJ. The experiment was performed at $T=0.3$ K on JJs with $I_c \simeq 2~\mu$A, $C\simeq 19$~pF, $\omega_{p0}/2\pi = 1.54$~GHz. The quality factor $Q_0\simeq 15-20$ can be deduced from the shape of the resonant curve, allowing explicit estimation of $R_{MW} = Q_0/(\omega_{p0} C) \simeq 100~\Omega$. The onset of bifurcation at $\omega \simeq 0.89~\omega_{p0}$ (1.375 GHz) was observed at the input power $P_{in} = 63.1$ fW (-102 dBm). My simulations for the same parameters yield the MW amplitude at the onset of bifurcation, $I_{MW} = 21.2$ nA, which corresponds to the absorbed power, Eq. (6), $P_a \simeq 22.5$ fW (-106.5 dBm). The ratio of calculated $P_a$ and measured $P_{in}$, $\chi = 0.36$, represents the absorption efficiency, Eq. (7). It is only slightly smaller than the maximum value, $\chi=0.5$. This can be attributed to an imperfect impedance matching between $R_{MW} \simeq 100~\Omega $ and $R_0=50 ~\Omega$ of the coaxial cable. Thus, simulations are in good agreement with the experimental data \cite{Siddiqi_2005}.

\subsection*{Limitations}

As seen from Fig. \ref{fig:fig4} (b), for realistic parameters of Nb/AlOx/Nb tunnel junctions with $Q_0=100$, the limits of sensitivity and noise-equivalent power at $T=1$ K reach outstanding value of $S\simeq 5\times10^{12}$ (V/W). 
However, it comes at a certain expense. 

Firstly, although the achievable sensitivity would be more than sufficient for single photon resolution in the MW range, however SCD at the maximal sensitivity has $~50\%$ dark count rate and, therefore, does not work as a single photon detector. It is possible to reduce the dark count rate by reducing bias, but this comes at the expense of reduced sensitivity, as shown in Fig. \ref{fig:fig3}. Yet, the sensitivity is so high that a certain reduction may be tolerated in order to enable a robust single photon detection. 

Secondly, at the optimal sensitivity SCD is quite slow. Because of a large dark-count rate, the response signal can be obtained only by collecting a large statistical ensemble, which takes time. On a positive side, lock-in measurement can be performed over many bias periods to reduce noise \cite{Borodianskyi_2017,Hovhannisyan_2021}. In the sub-optimal regime with deterministic photon counting, a single switching event is sufficient and, therefore, it could be fast.  

Thirdly, the resonant SCD is a narrow-band detector with the maximum sensitivity at $\omega=\omega_p$, as shown in Fig. \ref{fig:fig2} (a). For given experimental settings, optimal sensitivity is achieved at a frequency given by Eq. (18). However, a JJ is easily tunable. The plasma frequency $\omega_{p0}$ can be varied from the maximum value to zero by applying a small magnetic field, as shown in Fig. \ref{fig:fig4} (d). 

Beyond the resonant region and in the overdamed case, $Q_0<1$, a JJ may still have a good and broad-band non-resonant response, as demonstrated in Fig. \ref{fig:fig3} (a) for the case $\omega\ll\omega_{p0}$. For higher frequencies, $\omega>\omega_{p0}$, the response is caused by a simple shift of the switching histograms according to Eq. (\ref{I0Bessel}).  
Expanding the latter for small amplitudes and using Eq. (8) we obtain, $I_c/I_{c0} \simeq 1-2R_{MW}\chi (e/\hbar\omega)^2 P_{in}$. Since $V=G(I)V_g$ and $S=\partial V/ \partial P_{in}=V_g (\partial G/\partial I) (\partial I_c/\partial P_{in})=V_g g (\partial I_c/\partial P_{in})$, 
we obtain for the maximum sensitivity.
\begin{equation}
    S\sim 2\chi V_g (g_{max}I_{c0})R_{MW} \frac{e^2}{(\hbar \omega)^2}, ~~~~~ (\omega >\omega_{p0}).
\end{equation}
Taking into account that $g_{max}I_{c0}\sim 200$ in the case of Fig. \ref{fig:fig1} (b), for $\chi=0.5$ we obtain $S\sim 4 \times 10^8$ (V/W) at $f=100$ GHz. This is comparable to the low-frequency non-resonant sensitivity, shown in Fig. \ref{fig:fig3} (a). With increasing frequency the non-resonant response, Eq. (19), decays $\propto \omega^{-2}$. 
Note, that junction response does not vanish at $\omega \gg \omega_{p0}$. Tunnel JJs can detect even high-energy photons in a broad range from infrared to X-rays \cite{Soragga_2024,Wilson_2000,Angloher_2000}. However, the detection mechanism in that case is completely different and is caused by non-equilibrium phenomena induced by depairing \cite{Kozorezov_2003}, and not by the Josephson phase dynamics, considered here.

There can be other classical or quantum limitations. 
For example, the quality factor can be reduced by additional high-frequency dissipation channels \cite{Krasnov_2023}, which are non-negligible when $R_{QP}>R_{MW}$. Therefore, to avoid overestimation of SCD performance, I restricted my analysis to $Q_0\lesssim 100$, well below achievable MW quality factors in tunnel JJs \cite{Aprili_2012}.
The macroscopic quantum tunneling sets the limit for reducing the width of switching histograms \cite{Martinis_1987,Grabert_1987,Martinis_1988,Devyatov_1986}. But this is probably not be the major limitation because the MQT crossover temperature is typically in the 10 mK range and should only marginally affect JJs at $T=1$ K. Since the distinction between classical and quantum worlds is often obscure, eventual limitations of SCD, beyond the ``classical" description provided here, could be an interesting subject for further investigations. 

To conclude, I provided a quantitative description of a stochastic switching current detector, based on an underdamped Josephson junction. It is shown that the performance of SCD is very sensitive to junction parameters (critical current, quality factor, capacitance, Josephson plasma frequency, shunting resistance, e.t.c.) and experimental conditions (magnetic field, temperature, frequency).
The key physical phenomenon that boosts SCD performance is a resonant activation at a Josephson plasma frequency. 
The responsivity can be increased by several orders of magnitude at the optimal condition, corresponding to the bifurcation-less case with the plasma resonance aligned with the mid-point of the switching current histogram. The ways of tuning the detector for achieving optimal operation are discussed. For realistic parameters of Nb/AlOx/Nb tunnel junctions with $Q_0=100$, the sensitivity 
at $T=1$ K can reach limiting value of $S\simeq 5\times10^{12}$ (V/W). 


\section*{Erratum \cite{Erratum_2025}}

The simulations presented in the article are correct, including the predicted ultimate detector sensitivity $S \simeq 5~10^{12}$ (V/W). However, the ultimate Noise-Equivalent Power (NEP) deduced from this number is incorrect. The mistake was caused by having ignored the internal statistical (telegraph) noise in the junction and considering only the external preamplifier noise.  

The switching current detector (SCD) is probabilistic. The SCD voltage response has an expectation value and a variance. The voltage uncertainty for a 1 s measurement is, 
\begin{equation}
\frac{\delta V}{V_g} \simeq \pm \sqrt{\frac{G (1-G)}{2f_b}},    
\end{equation}
where $G$ is the switching probability, $f_b$ is the bias frequency, and $V_g\sim 3$ mV is the sum-gap voltage. The nominator in Eq. (1) represents the Bernoulli variance for a single switching event, and the denominator represents the statistical averaging of $2f_b$ switching events at positive and negative bias maxima during the 1 s time interval. 

The distinction criteria of two noisy signals is, 
\begin{equation}
    \Delta V > \sqrt{\delta V_1^2+\delta V_0^2}\simeq \sqrt{\frac{G (1-G)}{f_b}}V_g.
\end{equation}
At optimal sensitivity, $G=0.5$, and for $f_b =100$ Hz, $\Delta V \simeq 150~\mu$V/Hz$^{1/2}$, is much larger than the preamplifier noise, $\sim$ nV/Hz$^{1/2}$. Therefore, the NEP The intrinsic noise would reach parity with the extrinsic noise only at $f_b\sim 10$ GHz. Therefore, the NEP of SCD, under conventional low bias frequency operation, is determined by the intrinsic statistical uncertainty, $NEP = \Delta V/S$. 
The correct values of the ultimate NEP are $\sim 30$ aW/Hz$^{1/2}$ for $f_b=100$ Hz and $\sim 1$ aW/Hz$^{1/2}$ for $f_b=100$ kHz.

The ultralow NEP may be achievable only in the phase-diffusion regime, where a junction sporadically switches in and out of the resistive state at an attempt frequency close to the Josephson plasma frequency, $f_p$ \cite{Kautz_1990}. The large value of $f_p \sim 100$ GHz leads to a small statistical uncertainty, $\Delta V \sim$nV/Hz$^{1/2}$. As shown in Figures 4 (b), (d) and (f), the sensitivity at the edge of the phase diffusion regime remains very high, $S\sim 2-5\times 10^{12}$ V/W. Combined with a low-noise preamplifier, this could probably enable $NEP \sim~$zW/Hz$^{1/2}$.   

\end{document}